\documentclass[twocolumn,superscriptaddress, prb]{revtex4}
\usepackage{amsfonts}
\usepackage{amssymb}
\usepackage{amsmath}
\usepackage{amsthm}
\usepackage{graphicx}
\usepackage{stackrel}
\usepackage{color}

\usepackage{hyperref}

\newcommand{\ket}[1]{|#1\rangle}

\begin{document}
\title{Thermalization of entanglement}

\author{Liangsheng Zhang}
\affiliation{Physics Department, Princeton University, Princeton, NJ 08544}

\author{Hyungwon Kim}
\affiliation{Physics Department, Princeton University, Princeton, NJ 08544}
\affiliation{Department of Physics and Astronomy, Rutgers University, Piscataway, NJ 08854}

\author{David A. Huse}
\affiliation{Physics Department, Princeton University, Princeton, NJ 08544}

\begin{abstract}
We explore the dynamics of the entanglement entropy near equilibrium in highly-entangled pure states of two quantum-chaotic spin chains undergoing unitary time evolution.  We examine the relaxation to equilibrium from initial states with either less or more entanglement entropy than the equilibrium value, as well as the dynamics of the spontaneous fluctuations of the entanglement that occur in equilibrium.  For the spin chain with a time-independent Hamiltonian and thus an extensive conserved energy, we find slow relaxation of the entanglement entropy near equilibration.  Such slow relaxation is absent in a Floquet spin chain with a Hamiltonian that is periodic in time and thus has no local conservation law.
Therefore, we argue that slow diffusive energy transport is responsible for the slow relaxation of the entanglement entropy in the Hamiltonian system.
\end{abstract}


\maketitle

\section{Introduction}
Quantum entanglement has recently been a central topic in theoretical physics.
Many aspects of the dynamics of entanglement have been recently studied, such as
ballistic spreading of the entanglement in integrable \cite{amico,Calabrese:2005,Chiara:2006,Hartman:2013} and nonintegrable \cite{Kim:2013,Liu:2014} systems,
logarithmic spreading in many-body localized systems \cite{Znidaric:2008,Bardarson:2012}, and sub-ballistic spreading due to quantum Griffiths effects \cite{VHA}.
In many of these examples, the entanglement spreads more rapidly than
conserved quantities
that must be transported by currents.

\begin{figure}
\includegraphics[width=1.0\linewidth]{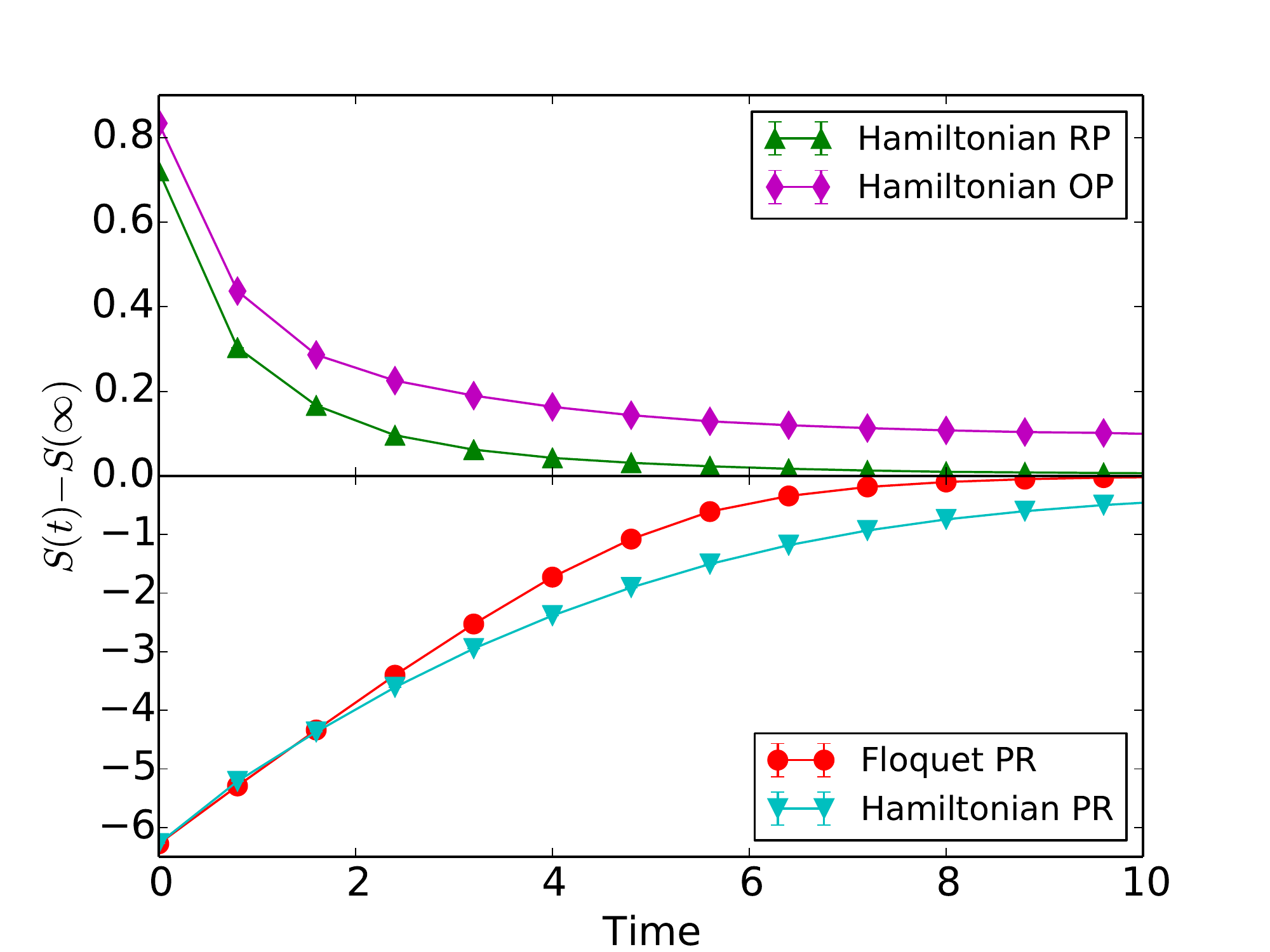}
\centering
\caption{(color online) Time evolution of the entanglement entropy for $L=14$ for: product random (PR) initial states under Floquet dynamics (2) with $\tau=0.8$ (red line with circles) as well as Hamiltonian dynamics (\ref{eqn:Hamiltonian}) (blue line with down triangles); ``generalized Bell'' initial states made from pairs of random pure (RP) states (green line with up triangles) and from ``oppositely paired'' (OP) states (purple line with diamonds) both under Hamiltonian dynamics.
Each case is averaged over 400 initial pure states, and the error estimates are too small to be visible in this figure.  See main text for more details.}
\label{fig:ent_evolution}
\end{figure}

Much of the previous work on the dynamics of entanglement, however, has emphasized far-from-equilibrium regimes,
particularly those following a quantum quench.
Here, we instead explore the entanglement dynamics near equilibrium in nonintegrable, thermalizing spin chains \cite{banuls} of finite length.
For example, if we start in a nonentangled initial pure state, the entanglement entropy grows linearly with time at early time
due to the ``ballistic'' spreading of entanglement \cite{Kim:2013,Liu:2014}, but then
saturates to its ``volume-law'' equilibrium value at long time. The lower two data sets in Fig.~\ref{fig:ent_evolution} illustrate this.
In the limit of a long spin chain, this isolated system is reservoir that thermalizes all of its subsystems.  Then the extensive part of the final equilibrium value of the entanglement entropy is equal to the thermal equilibrium entropy
at the corresponding temperature, and that temperature is set by the total energy of the initial state.
We call this process, in which the entanglement entropy approaches the thermal
equilibrium entropy, the ``thermalization of entanglement'' \cite{ETH}.

In this paper we focus on the late time, near equilibrium regime of the
entanglement dynamics, as well as the spontaneous fluctuations of entanglement in pure states sampled from the equilibrium density operator.
In section II, we introduce a nonintegrable, quantum-chaotic model Hamiltonian and its corresponding Floquet operator, where the extensive energy conservation
is removed.  In section III, we first examine the distribution of entanglement entropy of eigenstates of the Hamiltonian and the Floquet operators,
finding that the presence of the conservation law affects the distribution.  In section IV, we study the dynamics of entanglement entropy
near equilibrium.  We study three scenarios: starting from a product state of two random pure states, starting from generalized Bell states with
two different pairing schemes, and the autocorrelation of the spontaneous fluctuations of the entanglement entropy.  In all cases, we find the Floquet system
thermalizes entanglement faster than the Hamiltonian system.  In section V, we summarize our findings.

\section{Models}
To study a system that is robustly nonintegrable and strongly thermalizing, we choose the
spin-1/2 Ising chain
with both longitudinal and transverse fields.  Its Hamiltonian is
\begin{align}
\label{eqn:Hamiltonian}
H=\sum_{i=1}^L g\sigma_i^x + \sum_{i=1}^L h\sigma_i^z + \sum_{i=1}^{L-1}J\sigma_i^z\sigma_{i+1}^z ~,
\end{align}
where $\sigma_i^x$ and $\sigma_i^z$ are Pauli matrices at site $i$.
We use open boundary conditions and set the parameters to $(g,h,J) = (0.9045, 0.8090, 1.0)$,
for which this model has been shown to be robustly nonintegrable and strongly thermalizing for system sizes
readily accessible to exact diagonalization studies \cite{Kim:2013,KIH}.
The only conservation laws that this system is known to have at this parameter choice \cite{E8}
(other than projections on to its exact eigenstates) are
total energy, and parity under spatial reflection of the chain ($i\rightarrow (L+1-i)$).
This system's ``hydrodynamics'' are simply its conserved energy moving diffusively and subject to random local currents due to the system's
quantum-chaotic unitary dynamics.
We set the Planck constant $\hbar$ to unity
so that time and energy have inverse units of each other, and all energies and frequencies are in units of the interaction $J=1$.

To explore the effects of removing the conservation of total energy, we also study a Floquet system that is a modification of (\ref{eqn:Hamiltonian}).
We decompose the Hamiltonian into two parts, $H_z = \sum_i (h \sigma^z_i + \sigma^z_i \sigma^z_{i+1})$
and $H_x = \sum_i g \sigma^x_i$.  We periodically drive the system with a time-dependent Hamiltonian that is in turn $H(t)=2H_z$ for
a time interval of $\tau/2$
and then $H(t)=2H_x$ for the next $\tau/2$, and repeat.  The time-averaged Hamiltonian is thus unchanged, but the periodic switching changes the energy conservation
from conservation of the extensive total energy to conservation of energy only modulo $(2\pi/\tau)$.  This change removes the diffusive transport of energy as a slow
``hydrodynamic'' mode while otherwise changing the model as little as possible.
The Floquet operator that produces the unitary time evolution through one full period is
\begin{align}
U_F(\tau) = \mathrm{e}^{-i H_x \tau} \mathrm{e}^{-i H_z \tau} ~.
\end{align}
We choose time step $\tau = 0.8$, which was found in Ref. \cite{KIH} to produce
a rapid relaxation of the total energy within a few time steps as shown in the Appendix.
The eigenvalues of $U_F(\tau)$ are complex numbers of magnitude one.
Note that time is in a certain sense discrete (integer multiples of $\tau$) for this Floquet system.
The Hamiltonian system, with conserved total energy, is effectively the case $\tau=0$, which we contrast here with the Floquet system
with $\tau=0.8$ where the total energy is not conserved and relaxes very quickly.  Of course, there is an interesting crossover between these
two limits \cite{luca}, but we do not explore that crossover in this paper.

Throughout this paper,
we consider the bipartite entanglement entropy of pure states, quantified by the
von Neumann entropy of the reduced density operator of a
half chain: $S=-\mathrm{Tr}\{\rho_L \log_2\rho_L\}=-\mathrm{Tr}\{\rho_R \log_2\rho_R\}$.
We study chains of even length, and $\rho_L$ and $\rho_R$ are the reduced density operators of the left and right half chains, respectively.
Note that we measure the entropy in bits.

\section{Entanglement entropies of eigenstates}

\begin{figure}
\includegraphics[width=1.0\linewidth]{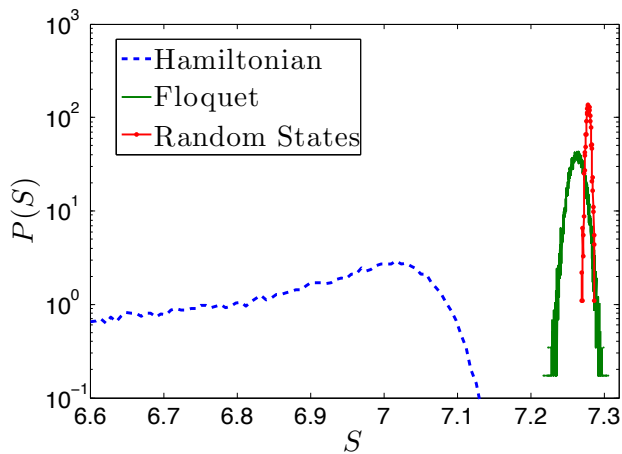}
\centering
\caption{(color online)  Normalized histogram of the entanglement entropy for $L=16$ for: (dashed blue line) the eigenstates of the Hamiltonian (\ref{eqn:Hamiltonian});
the eigenstates of the Floquet operator (Eq. (2)) with $\tau = 0.8$ (solid green line); and for random pure states of the full chain (very narrow red distribution), where the histogram is over 2000 randomly generated pure states. }
\label{fig:eigenstates}
\end{figure}

We first look at the entanglement entropy of the eigenstates of the Hamiltonian (1) and of the Floquet operator (2), compared to random pure states of the full chain.
Figure \ref{fig:eigenstates} shows the distributions of these entanglement entropies
for $L = 16$.

We can see that the entanglement of the eigenstates of the Floquet operator is close to
that of random pure states, first derived by Page \cite{Page:1993}:
\begin{equation}
S^R(L)=\frac{L}{2}-\frac{1}{2\ln 2}-\mathcal{O}\left(\frac{1}{2^L}\right) ~.
\label{eq:page}
\end{equation}
This is consistent with previous studies which have shown that a Floquet dynamics thermalizes a subsystem to infinite temperature \cite{luca, lazarides, ponte, KIH}.
The eigenstates of the Hamiltonian, on the other hand, all have entanglement entropies that are a fraction of a bit or more less than random pure states.
What is the source of this difference?  It is because the Hamiltonian eigenstates are eigenstates of the extensive conserved total energy, while the random
pure states and the Floquet eigenstates are not constrained by an extensive conserved quantity.  This causes the probability distribution of the energy of a
half chain to be narrower for the Hamiltonian system, since if one half chain has a high energy (compared to its share of the eigenenergy) then the other half chain
has to have an energy that is low by the same amount.  This suppresses the volume of the possible space
of half-chain states whose energy is either high or low, resulting in a reduced
entropy of the half-chain and thus reduced entanglement entropy, even for the Hamiltonian eigenstates at energies that correspond to infinite temperature.  This goes along with the recent observation that the finite-size deviations of the eigenstates of
the Hamiltonian from the Eigenstate Thermalization Hypothesis are larger than those of the eigenstates of the Floquet operator \cite{KIH}: energy conservation
somewhat impedes thermalization.

\section{Dynamics of Entanglement near equilibrium}
Now we turn to the dynamics of the entanglement entropy.
The dynamics of a linear operator is set by the matrix elements of the operator between
energy eigenstates (or eigenstates of the Floquet operator) and the eigenenergies.  But the entanglement is not a linear operator, so its dynamics cannot
be determined so simply.
We explore the near-equilibrium dynamics of the entanglement in two different ways. First we study the relaxation of the entanglement to its equilibrium value from particular initial states with either low or high entanglement. We then explore the dynamics of the spontaneous
fluctuations of the entanglement entropy during the unitary time evolution of a random pure state of the full spin chain.
From these studies we can clearly show that the entanglement dynamics is slower for the Hamiltonian system, since some of the entanglement entropy is connected to the slow diffusion of energy between the two half-chains.  In the Floquet system, on the other hand, near equilibrium the entanglement relaxes to equilibrium with a simple-exponential behavior in time, with a relaxation time that is apparently independent of the system size.

\subsection{Product of Random Pure states}

For initial states with zero entanglement between the two half chains, we use
a product of random (PR in Fig. 1) half-chain pure states $\ket{\psi(t=0)}=\ket{\psi_L}\otimes\ket{\psi_R}$,
where $\ket{\psi_L}$ and $\ket{\psi_R}$ are picked from the ensemble of random pure states of the left and the right half chain, respectively.
On average these states have energy close to 0, so the system is near infinite temperature and starts with zero entanglement entropy.
As random pure states are chosen for the half chains, the expected energy distribution between left and right halves is close to the equilibrium distribution (at infinite temperature), indicating only a small energy transport between two halves is required for thermalization.
Fig.~\ref{fig:ent_evolution} plots the time-dependent entanglement entropy under Hamiltonian and Floquet dynamics for $L=14$. The long-time average $S(\infty)$ is estimated by averaging $S(t)$ from $t=2500\tau$ to $t=2999\tau$.

It is clear from Fig.~\ref{fig:ent_evolution} that the Floquet system has faster relaxation of the entanglement entropy towards its saturation value at long times, even though the initial spreading rate of the entanglement is the same for these two systems.
Since the only significant difference between these two unitary dynamics is whether or not energy conservation and thus energy transport is present,
Fig.~\ref{fig:ent_evolution} suggests that the slow dynamical modes of this system associated with energy transport do also slow down the long-time thermalization of the entanglement.

\subsection{Generalized Bell States}

To explore the thermalization of the entanglement from initial states with {\it higher} entanglement than equilibrium, we use initial states
that maximize the entanglement entropy; we call these ``generalized Bell states''. These states have
Schmidt decomposition
\begin{equation}
\label{eqn:Bell}
\ket{\psi_B} = \frac{1}{\sqrt{2^{L/2}}}\sum_{i=1}^{2^{L/2}}\ket{L_i}\otimes\ket{R_i} ~,
\end{equation}
where the sets $\{\ket{L_i}\}$ and $\{\ket{R_i}\}$ are respectively complete orthonormal bases for left and right half chains.  Since these initial states have higher
entanglement entropy than equilibrium, their entropy {\it decreases} as it thermalizes.  This is an amusing apparent ``violation'' of the second law of thermodynamics,
but it is actually not thermodynamics, since the decrease is by less than one bit (very close to $1/(2\ln 2)$ by Eq. \eqref{eq:page})
, and thus far from extensive.

The random pure (RP) Bell states are made by independently choosing a random orthonormal basis for each half-chain.
To make initial Bell states that also have very large energy differences between the two half-chains, we make the opposite paired (OP) states that
can be written as
\begin{equation}
\ket{\psi(t=0)}=\frac{1}{\sqrt{2^{L/2}}}\sum_{i=1}^{2^{L/2}}\mathrm{e}^{i\theta_i}\ket{E_i^{h}}_{left}\otimes\ket{E_{2^{L/2}+1-i}^{h}}_{right} ~,
\end{equation}
where the $\ket{E_i^{h}}$ are the eigenstates of the half-chain Hamiltonian (Hamiltonian (\ref{eqn:Hamiltonian}) with $L/2$ sites), with their eigenenergies ordered according to $E_i^{h}\leq E_{i+1}^{h}$.
Therefore, by construction, many Schmidt pairs in these opposite paired Bell states have large energy imbalance between the two half-chains, unlike the random pure Bell states where the energy imbalance between the two halves is small.  The contrast between them shows how the slow diffusive relaxation of the energy imbalance affects entanglement thermalization.
The ensemble of OP states that we average over is obtained by choosing random phases $\{\theta_i\}$.

The time evolution of the entanglement entropy for a $L=14$ spin chain starting from generalized Bell states of pairs of random pure (RP) states as well as generalized Bell states with opposite pairing (OP) under Hamiltonian dynamics are shown in Fig.~\ref{fig:ent_evolution}, with the estimated long time average subtracted.
For the opposite paired (OP) initial states, the initial large energy differences between the two half-chains in many of the Schmidt pairs make the excess
entanglement long-lived, since the relaxation of these energy differences requires diffusion of the energy over the full length of the chain.
For the RP initial states, on the other hand, the half-chain states are random so do not show nonequilibrium energy correlations, and the excess entanglement
relaxes to equilibrium much
more rapidly than it does for the OP states.
This observation hence provides additional evidence of the coupling between entanglement entropy relaxation and energy transport under Hamiltonian dynamics.

\begin{figure}
\begin{center}
\includegraphics[width=1.0\linewidth]{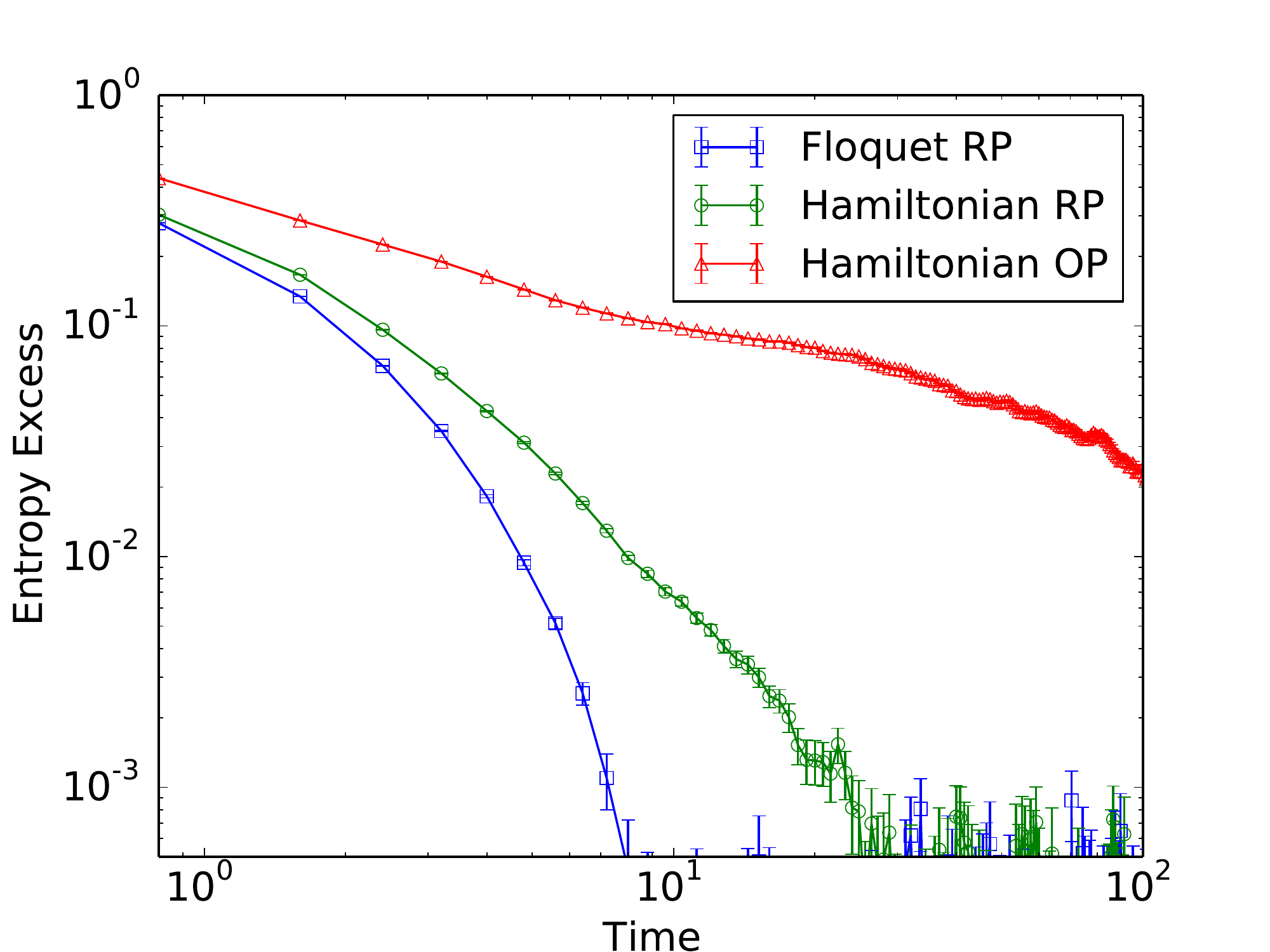}
\caption{(color online) Thermalization of
entanglement entropy in three cases for $L=14$ (log-log scale): from random pure (RP) generalized Bell states under Hamiltonian dynamics (circular markers, green);
from RP generalized Bell states under Floquet dynamics (square markers, blue); and from oppositely paired (OP) generalized Bell states under Hamiltonian dynamics
(triangular markers, red).  Within Hamiltonian dynamics, the larger initial energy imbalance for the OP initial states dictates a slower thermalization of the
entanglement, while the absence
of energy conservation for the Floquet system allows the fastest thermalization of entanglement entropy among all cases considered.
See main text for description of initial states.}
\label{fig:excess entropy}
\end{center}
\end{figure}

Fig.~\ref{fig:excess entropy} gives a more detailed view of the thermalization of the excess entanglement entropy starting from these generalized Bell initial
states. Here RP initial states under Floquet dynamics are also shown; since the Floquet system does not have conserved energy we cannot construct an OP
initial state for it.  This figure again shows the clear importance of energy transport for entanglement thermalization.
The excess entropy of the RP initial states decays away faster for the Floquet system as compared to the Hamiltonian system, since the thermalization of the Floquet system is not constrained by an extensive conserved energy.
The strong initial anticorrelation between the energies of
the two half-chains greatly slows down the thermalization of the entanglement for the OP initial states under Hamiltonian dynamics.

\subsection{Autocorrelation of entanglement}

Next we examine the dynamics of the spontaneous fluctuations of the entanglement entropy at equilibrium at infinite temperature, where
all pure states are equally likely.
Therefore, we simply pick many random pure states of the full chain and calculate the unitary time evolution of each initial state over many time steps.
We measure the autocorrelation of the entropy for each realization (indexed by $i$) as
\begin{equation}
\label{eqn:auto run}
R_i(t)=\frac{1}{M}\sum_{m=1}^{M}\left[S_i(t_m)-\bar{S}_i\right]\left[S_i(t_m+t)-\bar{S}_i\right] ~,
\end{equation}
where each run has in total 30000 time points $t_m$, equally spaced in time by $\Delta t$, and $S_i(T)$ is the entropy at time $T$.
Thus we measure the autocorrelation at integer multiples of the time step: $t = n\Delta t$.
$M$ is the maximum number of pairs that can be extracted from the time series.
Each random initial state
gives slightly different time-averaged entropies $\bar{S}_i$,
and thus for each run
we subtract its average in Eq. \eqref{eqn:auto run}.  Then we average over runs and normalize the autocorrelation to be one at time difference $t=0$.  The statistical errors are estimated from this averaging over runs.

\begin{figure}
\includegraphics[width=1.0\linewidth]{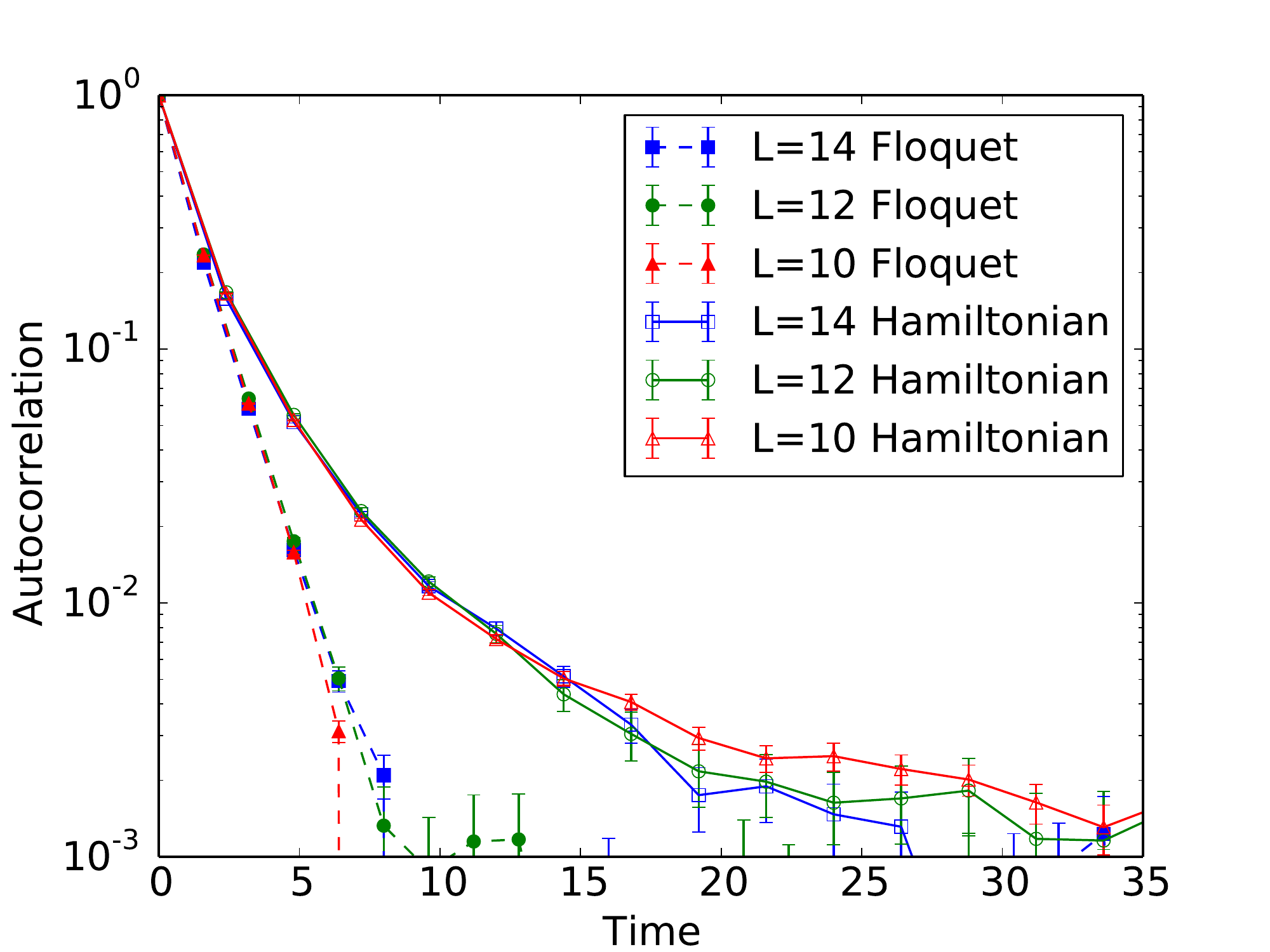}
\centering
\caption{(color online) Autocorrelation of the entanglement entropy vs. time for random pure states under Hamiltonian (solid lines) as well as Floquet dynamics (dashed lines) with different system sizes $L$ in log-linear scale.  For each case, the autocorrelation is normalized to be one at time zero.  Under Floquet dynamics, the autocorrelation decays as a simple exponential function of time, and faster than under Hamiltonian dynamics.  Only weak size dependence of this normalized autocorrelation is observed under either dynamics.}
\label{fig:autocorrelation}
\end{figure}

\begin{figure}
\includegraphics[width=1.0\linewidth]{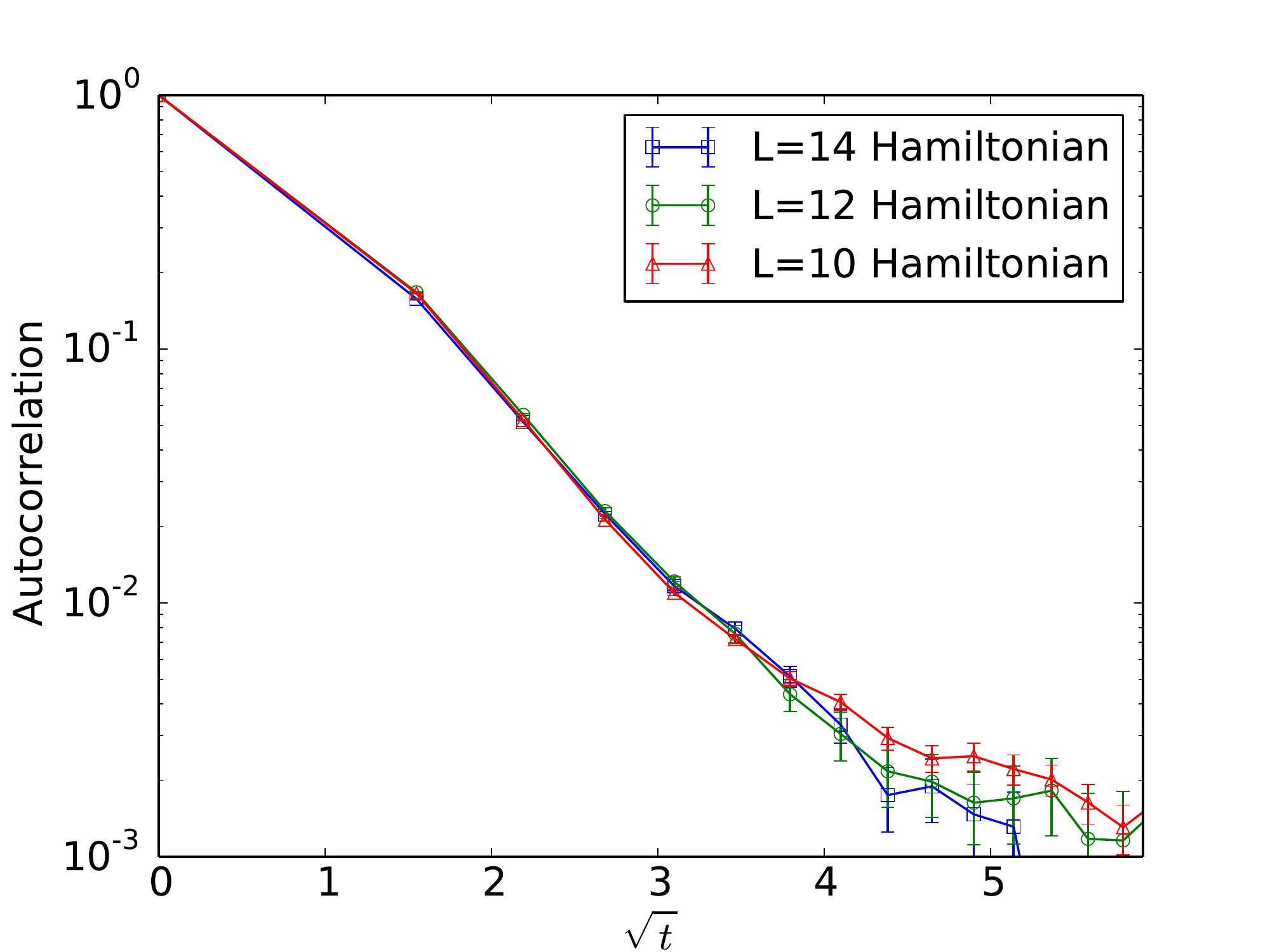}
\centering
\caption{(color online) Autocorrelation of the entanglement entropy under Hamiltonian dynamics vs. square root of time for random pure states with different system sizes $L$. On this semi-log scale, all three curves roughly follow a straight line and the tail becomes more straight as system size increases.}
\label{fig:Ham_auto}
\end{figure}

Fig.~\ref{fig:autocorrelation} plots the autocorrelation under Hamiltonian and Floquet dynamics with systems of different sizes: $L=10$, $L=12$ and $L=14$.
For $L=10$ the number of independent runs in each case is $N=400$, while for $L=12$ and $L=14$ we chose $N=100$.
With $\tau=0.8$ as before, our time points are spaced by
$\Delta t=3\tau$ for Hamiltonian dynamics and $\Delta t=2\tau$ for Floquet dynamics.
It can be easily seen from Fig.~\ref{fig:autocorrelation} that
the relaxation of autocorrelations under Floquet dynamics is systematically faster.
Particularly, the autocorrelation in the Floquet system assumes a simple exponential decay.
This observation indicates that under the Floquet dynamics a random state ``relaxes'' to equilibrium by independent and unconstrained local relaxation.
In the Hamiltonian system, on the other hand, a spontaneous fluctuation that rearranges the energy density on a long length scale is
necessarily slow, due to the slow energy diffusion.  Thus any influence of such fluctuations on the entanglement must relax slowly.
Clearly we are seeing such an influence that is causing the slower long-time relaxation of the entanglement autocorrelation in the Hamiltonian system.
Fig.~\ref{fig:Ham_auto} suggests that the autocorrelation under Hamiltonian dynamics decays exponentially in square root of time, as curves become roughly straight when plotted against $\sqrt{t}$ in a semi-log plot, and the straightness increases as system size increases. This scaling may be understood as fluctuation of entanglement entropy coupled to operators on the spin chain.
At time $t$ the fluctuation couples to the $O(4^l)$ operators on a size of $l\sim\sqrt{t}$ by diffusion, of which only $O(1)$ operators are slow, so only $O(1/4^l)\sim\exp(-c\sqrt{t})$ fraction of the information about the initial fluctuation is left at time $t$, where $c$ is some constant, resulting in an exponential decay of autocorrelation in $\sqrt{t}$.
The same reasoning may also be applied to the Floquet system, where the slowest modes instead have \cite{Kim: 2014} $l\sim t$, thus leading to a the observed simple exponential decay as is shown in Figure \ref{fig:autocorrelation}.

One may also note here that under either dynamics, the relaxation of these autocorrelations has little dependence on system size.
This indicates that the fluctuations that are contributing here are on length scales smaller than the $L=10$ systems.
For the Floquet system this is consistent with the relaxation being simply local, so any longer length scale slow operators \cite{Kim: 2014}
apparently do not couple substantially to the entanglement fluctuations.
For the Hamiltonian systems this absence of size dependence suggests that over the time range probed here, the energy fluctuations that couple to the
entanglement are on length scales smaller than the length of the smaller $L=10$ system.  But the substantially slower relaxation as compared to the
Floquet system suggests that energy transport over a few lattice spacings does couple to the entanglement fluctuations.

\section{Conclusion}

In conclusion, we have investigated the thermalization of the entanglement entropy by comparing state evolution of spin chains under Hamiltonian dynamics and Floquet dynamics, with the two systems having the same time-averaged Hamiltonian.
Eigenstates of these two dynamics have quite different distributions of the entanglement entropy.  The Floquet eigenstates all have entanglement close to that of random pure states, while the Hamiltonian eigenstates all have significantly less entanglement due to the constraint of total energy conservation.
We show that the entanglement entropy relaxes to equilibrium more slowly under Hamiltonian dynamics, both for initial states well away from equilibrium and for the
spontaneous fluctuations of the entanglement entropy at equilibrium.  The Hamiltonian system has slow diffusive energy transport, while the Floquet system does not.
This slow diffusive relaxation of the energy distribution in the Hamiltonian system results in slow relaxation near equilibrium of the entanglement entropy.

\section{Acknowledgement}
H.K. is supported by NSF DMR-1308141.

\appendix*

\newpage

\section*{Appendix}
\begin{figure}
\begin{center}
\includegraphics[width=1.0\linewidth]{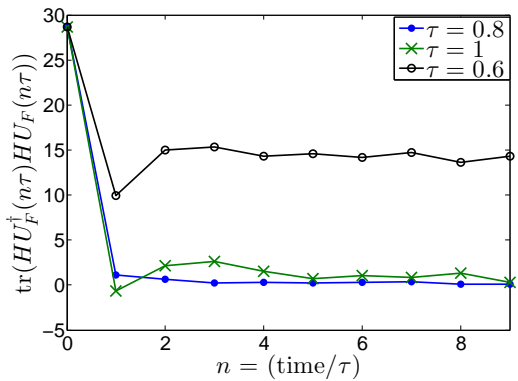}
\caption{(color online) Relaxation of total energy under Floquet dynamics for $L=12$ with different choices of driving period $\tau$.}
\label{fig:tau choice}
\end{center}
\end{figure}

To choose the time step $\tau$ for our Floquet system, we investigated the relaxation of the total energy under different choices of $\tau$ for $L=12$, as is shown in Fig.~\ref{fig:tau choice}.
Particularly, it demonstrates how the autocorrelation of the total energy decays under discrete Floquet dynamics: $\mathrm{Tr}\{HU^{\dagger}_F(n\tau)HU_F(n\tau)\}$. 
Both $\tau=0.8$ and $\tau=1$ relax the total energy very quickly so that the system approaches infinite temperature within of order one time step, while for $\tau=0.6$ the energy relaxation has a much slower component.
$\tau=1$ leads to a more oscillatory behaviour compared with $\tau=0.8$.  Therefore $\tau=0.8$ emerges as a nice choice for our studies and those of Refs.\cite{KIH,Kim: 2014}.

\end{document}